\documentclass{article}
\usepackage{amsmath}

\begin{document}

\title{Modeling of temperature and excitation dependences of efficiency in
an InGaN light-emitting diode }
\author{Weng W. Chow \\
Sandia National Laboratories,\\
Albuquerque, New Mexico 87185-1086, USA \\
wwchow@sandia.gov\\
\noindent\ }
\maketitle

\begin{abstract}
The changes in excitation dependence of efficiency with temperature is
modeled for a wurtzite InGaN light-emitting diode. \ The model incorporates
bandstructure changes with carrier density arising from screening of
quantum-confined Stark effect. Bandstructure is computed by solving Poisson
and k$\cdot $p equations in the envelop approximation. The information is
used in a dynamical model for populations in momentum-resolved electron and
hole states. Application of the approach shows the interplay of quantum-well
and barrier emissions giving rise to shape changes in efficiency versus
current density with changing temperature, as observed in some experiments.

\bigskip 
\noindent OCIS: {(230.3670) Light-emitting diodes; (230.5590)
quantum-well, -wire and -dot devices; (250.5590)
quantum-well, -wire and -dot devices.}
\end{abstract}


\section{Introduction}

Considerable resources are being devoted to advancing InGaN light-emitting
diodes (LEDs), largely because of solid-state lighting. Of particular interest is
improving device efficiency \cite{krames}, especially the mitigation of efficiency
degradation with increasing excitation, i.e., efficiency droop \cite%
{shen,hader2,dellaney}. To arrive at a solution, it is important to understand
the principal mechanism of emission together with the effects of
carrier leakage \cite{schubert}, Auger recombination \cite{shen}, junction
heating \cite{efremov}, carrier and defect delocalizations \cite%
{chichibu,smowton,hader1}. Equally important is the change in efficiency with
temperature, not only for application but also for better
understanding of the physics contributing to efficiency droop.

This paper investigates LED\ efficiency as functions of current density and
lattice temperature. The analysis uses a model that allows direct input of
band-structure properties \cite{chow1,chow2}. Band-structure details are
important, because underlying emission properties in wurtzite quantum-well
(QW) structures are the changes in band energy dispersions, confinement
energies and optical transition matrix elements with excitation. These
changes arise from screening of piezoelectric and spontaneous polarization
fields \cite{shur,hangleiter,kira}.

The present approach differs substantially from the total-carrier-density
rate-equation description used in most discussions involving InGaN LED
efficiency. In terms of reproducing experimental efficiency versus injection
current data, a particularly successful rate-equation model is the $ABC$
model \cite{shen,ryu}. The model's name derives from the phenomenological
constants $A$, $B$ and $C,$ introduced to account for Shockley-Read-Hall
(SRH), radiative-recombination and Auger-scattering carrier losses,
respectively. Band-structure properties enter indirectly via these
coefficients.

Section 2 summarizes the model and presents the working equations.\ Section
3 demonstrates the application of the model by calculating internal quantum
efficiency (IQE) as a function of injection current for different lattice
temperatures. Following Ref. \cite%
{hader3}, the calculations are performed for a LED\ with an In$_{0.37}$Ga%
$_{0.63}$N QW active region. The choices of input parameters are explained
and comparison to recent experimental results is discussed. This section
also describes the contributions from Auger carrier loss, plasma heating and
carrier leakage. Then, the role of the band structure is illustrated by
considering a LED\ with an In$_{0.20}$Ga$_{0.80}$N active region. Section 4
summarizes the paper.

\section{InGaN LED model}

The derivation of spontaneous emission from QW and barrier transitions
starts with a Hamiltonian adapted from quantum optics \cite{jaynes}. With
this Hamiltonian, Hiesenberg operator equations of motion are derived for
the carrier populations and polarizations. These equations contained
operator products describing coupling to higher order correlations in
light-matter interaction. Truncating at the Hartree-Fock level, as well as
adding phenomenologically SRH carrier loss and relaxation contributions from
carrier-carrier and carrier-phonon scattering, lead to the working equations
for the present investigation \cite{chow2}.

For the QW population $n_{\sigma ,\alpha _{\sigma },k_{\bot }}$,
\begin{eqnarray}
\frac{\text{d}n_{\sigma ,\alpha _{\sigma },k_{\bot }}}{\text{d}t}
&=&-n_{\sigma ,n_{\sigma },k_{\bot }}\sum\limits_{\alpha _{\sigma ^{\prime
}}}b_{\alpha _{\sigma },\alpha _{\sigma ^{\prime }},k_{\bot }}n_{\sigma
^{\prime },\alpha _{\sigma ^{\prime }},k_{\perp }}-An_{\sigma ,n_{\sigma
},k_{\bot }}  \notag \\
&&-\gamma _{c-c}\left[ n_{\sigma ,n_{\sigma },k_{\bot }}-f\left( \varepsilon
_{\sigma ,k_{\perp }},\mu _{\sigma },T\right) \right]  \notag \\
&&-\gamma _{c-p}\left[ n_{\sigma ,n_{\sigma },k_{\bot }}-f\left( \varepsilon
_{\sigma ,k_{\perp }},\mu _{\sigma }^{L},T_{L}\right) \right]  \notag \\
&&-\gamma _{ag}\left[ n_{\sigma ,n_{\sigma },k_{\bot }}-f\left( \varepsilon
_{\sigma ,k_{\perp }},\mu _{ag},T_{ag}\right) \right]  \label{e1}
\end{eqnarray}%
where each QW state is denoted by the subscripts $\sigma ,\alpha
_{\sigma },k_{\bot }$, representing charge ($\sigma =e$ or $h$), subband
index ($\alpha _{\sigma }$) and in-plane momentum ($k_{\perp }$). In the
above equation, $A$ is the SRH coefficient, $\gamma _{c-c}$ and $\gamma
_{c-p}$ are the effective carrier-carrier and carrier-phonon collision
rates, respectively. Carrier loss by Auger scattering is modeled with a term
containing an effective Auger scattering rate $\gamma _{ag}=CN^{2}$, where $%
C $ is the Auger coefficient and $N$ is the total (QW and barrier) carrier
density. In the summation, $\sigma ^{\prime }=h$ or $e$, when $\sigma =e$ or 
$h$.

For the barrier carrier population $n_{\sigma ,k}^{b}$,
\begin{eqnarray}
\frac{\text{d}n_{\sigma ,k}^{b}}{\text{d}t} &=&-b_{k}n_{e,k}^{b}n_{h,k}^{b}+%
\frac{J}{eN_{\sigma }^{p}}f\left( \varepsilon _{\sigma ,k}^{b},\mu _{\sigma
}^{p},T_{p}\right) \left( 1-n_{\sigma ,k}^{b}\right) -An_{\sigma ,k}  \notag
\\
&&-\gamma _{c-c}\left[ n_{\sigma ,k}^{b}-f\left( \varepsilon _{\sigma
,k}^{b},\mu _{\sigma },T\right) \right]  \notag \\
&&-\gamma _{c-p}\left[ n_{\sigma ,k}^{b}-f\left( \varepsilon _{\sigma
,k}^{b},\mu _{\sigma }^{L},T_{L}\right) \right]  \notag \\
&&-\gamma _{ag}\left[ n_{\sigma ,k}^{b}-f\left( \varepsilon _{\sigma
,k}^{b},\mu _{ag},T_{ag}\right) \right]  \label{e2}
\end{eqnarray}%
where each barrier state is denoted by the subscripts $\sigma ,k$,
representing charge ($\sigma $) and 3-dimensional momentum ($k$). In the
above equation, $J$ is the current density, $e$ is the electron charge and $%
N_{\sigma }^{p}=\sum_{k}f\left( \varepsilon _{\sigma ,k}^{b},\mu _{\sigma
}^{p},T_{p}\right) $. The injected carrier distribution $f\left( \varepsilon
_{\sigma ,k}^{b},\mu _{\sigma }^{p},T_{p}\right) $ is a Fermi-Dirac function
with chemical potential $\mu _{\sigma }^{p}$, temperature $T_{p}$.

In the above equations, $\varepsilon _{\sigma ,k_{\perp }}$ and $\varepsilon
_{\sigma ,k}^{b}$ are the carrier energies in the QW and barrier, respectively.
From the derivation, the spontaneous emission coefficients are given by
\begin{equation}
b_{\alpha _{\sigma },\alpha _{\sigma ^{\prime }},k_{\bot }}=\frac{1}{\hbar
\epsilon _{b}\pi c^{3}}\left\vert \wp _{\alpha _{\sigma },\alpha _{\sigma
^{\prime }},k_{\bot }}\right\vert ^{2}\Omega _{\alpha _{\sigma },\alpha
_{\sigma ^{\prime }},k_{\bot }}^{3}  \label{e3}
\end{equation}

\bigskip 
\begin{equation}
b_{k}=\frac{1}{\hbar \epsilon _{b}\pi c^{3}}\left\vert \wp _{k}\right\vert
^{2}\Omega _{k}^{3},  \label{e4}
\end{equation}%
where $\wp _{\alpha _{\sigma },\alpha _{\sigma ^{\prime }},k_{\bot }}$ and $%
\Omega _{\alpha _{\sigma },\alpha _{\sigma ^{\prime }},k_{\bot }}$ are the
QW dipole matrix element and transition energy, $\wp _{k}$ and $\Omega _{k}$
are the barrier dipole matrix element and transition energy, and $\epsilon
_{b}$ is the host permittivity.

For the asymptotic Fermi-Dirac distributions approached via carrier-carrier
collisions $f\left( \varepsilon _{\sigma ,k_{\perp }},\mu _{\sigma
},T\right) $ and $f\left( \varepsilon _{\sigma ,k}^{b},\mu _{\sigma
},T\right) $, the chemical potential $\mu _{\sigma }$ and plasma temperature 
$T$ are determined by conservation of carrier density and energy. For $%
f\left( \varepsilon _{\sigma ,k_{\perp }},\mu _{\sigma }^{L},T_{L}\right) $
and $f\left( \varepsilon _{\sigma ,k}^{b},\mu _{\sigma }^{L},T_{L}\right) $,
which are reached via carrier-phonon collisions, the chemical potential $\mu
_{\sigma }^{L}$ is determined by conservation of carrier density and the
lattice temperature $T_{L}$ is an input quantity. For $f\left( \varepsilon
_{\sigma ,k_{\perp }},%
\mu
_{ag},T_{ag}\right) $ and $f\left( \varepsilon _{\sigma ,k}^{b},%
\mu
_{ag},T_{ag}\right) $, which are reached via Auger scattering, there is a
loss of carrier density that eventually leads to a chemical potential $%
\mu
_{ag}$ of the intrinsic semiconductor, i.e., approximately at mid gap, with
its exact location depending on electron and hole dispersions. $T_{ag}$ is
determined by noting that energy is conserved in an Auger scattering event.

When summed over all states, Eqs. (1) plus (2) reduce to total carrier
density rate equations used in the $ABC$ model, if we assume that the total
emission contribution equals $BN^{2}$, where $B$ is an effective bimolecular
radiative coefficient. In the summation over states, quantities such as the
total carrier density and energy are computed by converting the sum over
states to integrals, i.e., 
\begin{equation}
\sum\limits_{k_{\perp }}\rightarrow \frac{S}{\left( 2\pi \right) ^{2}}%
2\int\limits_{0}^{\infty }\text{d}k_{\perp }\ 2\pi k_{\perp }\text{ \ and \ }%
\sum\limits_{k}\rightarrow \frac{hS}{\left( 2\pi \right) ^{3}}%
2\int\limits_{0}^{\infty }\text{d}k\ 4\pi k^{2}  \label{e5}
\end{equation}%
where $S$ and $h$ are the surface area and thickness of the active active
region consisting of the QW and barrier layers. The conversion to bulk (3-d)
density is via division by the QW layer width $h$ in the case of the QW and
by the total barrier width $h_{b}$ in the case of the barrier. Further
discussions involving implementation and comparison with results from
quantum-kinetic calculations may be found in earlier reports \cite%
{chow1,chow2,waldmueller}.\ The dynamical solution gives the carrier
densities in QW and barrier states.

Band-structure information enters directly into Eqs (\ref{e1}) and (\ref{e2}%
) via the dipole matrix elements $\wp _{\alpha _{\sigma },\alpha _{\sigma
^{\prime }},k_{\bot }}$, $\wp _{k}$ and carrier energies $\varepsilon
_{\sigma ,k_{\perp }}$, $\varepsilon _{\sigma ,k}^{b}$. Iterative solution
of the $k\cdot p$ and Poisson equations \cite{chuang} gives the QW energies
and dipole matrix elements. For this calculation, the necessary bulk
wurtzite material parameters are given in Refs. \cite%
{jenkins,wright,wei,ambacher}. When performing the band-structure
calculation, quasiequilibrium condition is assumed to determine the QW and
barrier bulk densities used in the solution of Poisson equation. This is an
inconsistency that is acceptable provided the dynamical solution does not
produce carrier distributions deviating too far from quasiequilibrium
distributions.

Application of the model involves first calculating the band structure for a
range of carrier densities by solving the $k\cdot p$ and Poisson equations.
This gives the band-structure information needed for the solving the
population dynamical equations. To facilitate the numerics, carrier states
are grouped into two categories: those belonging to the QWs and those
belonging to the barriers. With a high internal electric field, the
distinction between QW and barrier states may be ambiguous. In this paper,
the choice is made by calculating $\int_{QW}$dz$\ \left\vert u_{m_{\sigma
},\beta _{\sigma }}(z)\right\vert ^{2}$, where integral is performed over
the QWs and $u_{m_{\sigma },\beta _{\sigma }}$ is the wavefunction. The
states where the integral is greater than $1/2$ are grouped as QW states
and the rest as barrier states. For the problem being addressed, which are
the excitation and temperature dependences of IQE, the distinction is only
important because only QW transitions are affected by the quantum-confined
Stark effect (QCSE). For the barrier
transitions, the dipole matrix element in the presence of an internal
electric field is approximated by an average, where each transition is
weighted according to the occupations of the participating states. Grouping
the barrier states appreciably reduces numerical demand, which remains
substantial because one is still keeping track of a large number of $k$%
-states.

In the numerical solution of Eqs. (\ref{e1}) and (\ref{e2}), the
band-structure quantities are updated at each time step according to the
instantaneous carrier density. When steady state is reached, IQE is
determined from dividing the rate of carrier (electron or hole) loss via
spontaneous emission by the rate of carrier injection, i.e.
\begin{equation}
IQE=\frac{e}{JS}\left( \sum\limits_{\alpha _{e},\alpha _{h},k_{\perp
}}b_{\alpha _{e},\alpha _{h},k_{\bot }}n_{e,\alpha _{e},k_{\perp
}}n_{h,\alpha _{h},k_{\perp
}}+\sum\limits_{k}b_{k}n_{e,k}^{b}n_{h,k}^{b}\right)  \label{e6}
\end{equation}

\section{Results}

The above model was applied to compute the temperature dependence of IQE
versus current density. A motivation is that adding temperature dependence
may produce further physical insight to efficiency droop, as well as
provide more stringent testing of the model. In order to relate to experiments \cite%
{expt1,expt2,hader3}, simulations were performed for two LED\
configurations. The experimental devices had single-QW active regions, which
circumvented complications arising from nonuniform population in multi-QW
structures. One device consisted a 2nm In$_{0.37}$Ga$_{0.63}$N QW between
GaN barriers \cite{expt1}, while the other device consisted a 3nm In$_{0.20}$%
Ga$_{0.80}$N QW between GaN barriers \cite{expt2}.

The experiments \cite{expt1,expt2,hader1} show features of IQE dependences
on excitation and temperature, that if understood, may shed light on LED
physics in general and the efficiency droop in particular. They involve
changes in shape of IQE versus current density curves with temperature and
the relationships among these curves after the onset of droop. In the
former, for lattice temperatures $300K$ and above, both In$_{0.37}$Ga$%
_{0.63} $N and In$_{0.20}$Ga$_{0.80}$N LEDs exhibit IQE versus current
density behavior showing the familiar efficiency droop, as described
relatively well by the ABC model. Interestingly, a second IQE bump appears
at lower temperatures for the In$_{0.37}$Ga$_{0.63}$N sample and becomes
very pronounced at 100K. The ABC model is unable to describe this shape
change, which does not occur in the In$_{0.20}$Ga$_{0.80}$N sample at any
temperature. A goal of this study is to use our model to reproduce and
understand the origin of the second bump and determine whether the different 
behavior between the two LEDs is intrinsic, i.e., involving only differences in
band structure. Additionally, the model will be used to examine the detailed
relationship among IQE curves at high current densities. In the In$_{0.37}$Ga%
$_{0.63}$N case, the curves for different temperatures are clearly separated
for the entire measured current density range. On the other hand, for In$%
_{0.20}$Ga$_{0.80}$N, the IQE versus current curves for different
temperatures either cross or merge with each other at
high current densities. Previous modeling of these behaviors has lead to
puzzling results, such as inferring a decrease in $C$ coefficient with
increasing temperature, contrary to what is expected of Auger scattering 
\cite{hader1}.

Figure 1 shows results from the present model for the In$_{0.37}$Ga$_{0.63}$%
N LED. \ The curves describe IQE\ versus current density at different
lattice temperatures. Similar to experiment, there is a double bump
excitation dependence in IQE. The lower excitation bump decreases with
increasing temperature and vanishes for $T_{L}>250K$. Input parameters that
are assumed temperature dependent are the SRH and Auger coefficients, $A$
and $C$, respectively and the carrier-phonon scattering rate $\gamma _{c-p}.$
They are adjusted to produce IQE behavior resembling those found in Refs. 
\cite{expt1} and \cite{hader3}. The SRH coefficient is decreased with
decreasing temperature to give the onset of emission as depicted in the
figure. The Auger coefficient is adjusted to reproduce approximately the
experimental maximum IQE values. Since Auger carrier loss conserves energy,
significant plasma heating may occur. The role of carrier-phonon scattering
is to relax the hot carrier distributions back to the lattice temperature.
Both $C$ and $\gamma _{c-p}$ affect the peak IQE. However, the current
density where IQE peaks is relatively sensitive to $\gamma _{c-p}$, which is
chosen to produce an onset of droop in the tens of A/cm$^{2}$ range.

Figure 2 plots the values of $A$, $C$ and $\gamma _{c-p}$ versus temperature
used in computing the curves in Fig. 1. A decrease in $A$ with decreasing
temperature is expected for defect related loss. Figures 2 (b) and 2 (c)
show increases in Auger coefficient and carrier-phonon scattering rate with
increasing temperature, which are consistent with microscopic calculations.
The values of Auger coefficient are within the range predicted for
phonon-assisted Auger scattering \cite{dellaney} and smaller than most
values obtained from experimental curve fitting with the ABC model \cite{ryu}%
. The values used for carrier-phonon scattering are higher than predicted by
quantum kinetic calculations for typical III-N structures. However, it
should be noted that those calculations are for intraband scattering among
nearby states. The present effective rates represent the relaxation of very
energetic states populated by Auger scattering to the QW ground state. The
energy differences are in the neighborhood of the bandgap energy ($2.7eV$).

To give some insight into the origin of the double bump, Fig. 3 (a) shows
the spontaneous emission contributions from QW and barrier states versus
current density for the $T_{L}=200K$ curve in Fig. 1. Clearly, the spontaneous
 emission from barrier states dominates the emission from the 
QW at low current density, coinciding with the location of first IQE bump
[see left arrows in Figs. 1 and 3 (a)]. Here, optical emission from barrier
transitions occurs via the contribution $\sum_{k}b_{k}n_{e,k}^{b}n_{h,k}^{b}$%
, as soon as the product of electron and hole populations, $%
n_{e,k}^{b}n_{h,k}^{b}$ becomes nonzero. However, there may be no optical
emission from QW transitions because of absence of localized QW states from
the strong piezoelectric and spontaneous polarization fields
(quantum-confined Stark effect). There is a transition region where the
screening of internal electric field results in the appearance of localized
QW states. Here, the small energy separation between QW and barrier states
gives comparable QW and barrier populations, especially for the holes.
However, the QW emission contribution $\sum_{\alpha _{e},\alpha
_{h},k_{\perp }}b_{\alpha _{e},\alpha _{h},k_{\bot }}n_{e,\alpha
_{e},k_{\perp }}n_{h,\alpha _{h},k_{\perp }}$ can still be negligible, even
though the product $n_{e,\alpha _{e},k\bot }n_{h,\alpha _{h},k\bot }$ may be
appreciable, because QCSE spatially separates electrons and holes in the
QWs, resulting in very small dipole matrix elements for QW transitions. At
high current density [see right arrows in Fig. 1 and 3 (a)], plasma
screening restores the QW localized states and optical-dipole matrix
element, leading to QW emission overtaking that from the barrier because of
the advantage in 2-dimensional versus 3-dimensional carrier density of
states.

The above explanation is supported by Figs. 3 (b) and 3 (c). The solid
curves are absolute square of envelope functions for electrons and holes at
current densities indicated by arrows in Fig. 3 (a). Each curve is displaced
according to its energy for clarity. Envelope functions belonging to QW and
barrier states are indicated by black and grey curves, respectively. The
former is clearly missing in Fig. 3 (b). The black dashed lines plot the
confinement potentials and they indicate a significantly stronger
deformation in Fig. 3 (b) compared to Fig. 3 (c) due to the higher internal
electrical fields in the lower current density situation.

Figure 4 plots the contributions to the $T_{L}=200K$ IQE curve for various recombination
and scattering processes. The solid red curve is the sum of QW and barrier emission,
with a slight change in slope at low current density indicating the transition from
predominately barrier to predominately QW emission. A second slope change at
 higher current density arises from the onset of Auger carrier
loss. SRH and leakage (or failure to capture) losses from the barrier \ is
depicted by the blue dashed curve. They are primarily responsible for the
low IQE at low current densities. The green dotted curve shows that defect
(SRH)\ loss from QW states are essentially negligible. Auger carrier loss
accounts for the difference between the solid black curve and the sum of all
the other curves.

To illustrate the Auger contribution, Fig. 5 (a) shows the IQE versus
current density for two Auger coefficients: $C=2.3\times
10^{-31}cm^{6}s^{-1} $, which is used in Fig. 1, and $C=10^{-34}cm^{6}s^{-1}$%
, which is in the neighborhood of what is typically expected for materials
with bandgap energy of approximately 2.7eV. The dashed curve clearly shows
that for $C$ less than $10^{-34}cm^{6}s^{-1}$, Auger carrier loss has negligible
effect on IQE. The difference between the two curves indicates the Auger
contribution to the IQE curve in Fig. 1 for $T_{L}=200K$. During each Auger
event, an electron and a hole recombine, with the loss in energy transferred
to a second electron or hole. In a wide bandgap material, such as InGaN, the
transferred energy is significant and can lead to highly-energetic carrier
distributions, depending on the carrier-phonon scattering rate. Because of
rapid carrier-carrier scattering, this change in carrier distributions may
be described in terms of the plasma temperature $T$. Figure 5 (b) shows the
plasma temperature versus current density for the two Auger coefficients.
The solid curve indicates significant rise in temperature for the $T_{L}=200K$
curve in Fig. 1. The dashed curve deoicts the rise in temperature in the
absence of Auger loss, so that heating comes mainly from the relaxation of
carriers from the barrier states, where they are injected, to the ground
states, where emission occurs.

The role of band structure is investigated by performing simulations for the
In$_{0.20}$Ga$_{0.80}$N LED, using the same parameter values as plotted in
Fig. 2. Figure 6 shows the IQE versus current density for different
temperatures. Clearly indicated is a missing or negligible second IQE bump.
Band-structure calculations traced the reason to a smaller internal electric
field with lower In concentration in the QW. Consequently, localized QW
states exist for the entire temperature and current-density range
considered, and their emission exceeds that of the barrier states at all
current densities. The absence of a transition from barrier to QW emission
dominance gives rise to the lack of a double bump shape.

Experimental evidence supporting the above argument may be found in \cite%
{expt2}. There, the IQE versus current density curves shows single-bump
behavior for $T_{L}$ between $100K$ and $300K$, as in Fig. 6. However,
agreement is not complete for the simulations fail to exactly reproduce experimental
observation at high current density, where the experiment shows crossing of the IQE
curves for different temperatures.

At high excitation, the fitting parameters used in this study which
influence LED behavior are C and $\gamma _{c-p}$. We had assumed that
their values do not change considerably with a change in Indium concentration
 from 
Ref. \cite{expt2}, we again consider A, C and $\gamma _{c-p}$ as
free-parameters. Figure 7 shows the IQE results for input parameters given in
Fig. 8. Figure 7 (a) indicates better agreement with with Ref. 14. However,
agreement between theory and experiment for the In$_{0.37}$Ga$_{0.64}$N LED
with the new set of parameters is degraded. On the other hand the single-
versus double-bump behavior remains consistent with experiment and appears
robust to changes in the fitting parameters.

\section{Summary}

This paper uses an approach to modeling InGaN LEDs that involves the
self-consistent solution of band structure and carrier dynamics.
One motivation is systematic incorporation into a model, at a microscopic
level and on equal footing, the effects of spontaneous emission, carrier
capture and leakage, as well as nonequilibrium effects such as plasma heating. \
Another motivation is to provide direct input of band-structure properties,
in particular, their carrier-density dependences arising from screening of
piezoelectric and spontaneous polarization fields. Doing so allows
comparison of different devices at the heterostructure-design level. 

The end result is a non-equilibrium microscopic model that provides a more
precise estimation of relative strengths of possible physical contributions
compared to the commonly used $ABC$ model \cite{shen} and is easier to implement
than a first-principles, many-body approach \cite{hader2}. The model
reproduces the main experimental features of the observed droop behavior
versus temperature relatively successfully. Shape changes of IQE versus
current density with temperature are described with the optical emission
from quantum-well and barrier transitions treated on equal footing. The IQE
droop is described by Auger scattering and resulting plasma heating, with
material parameters and their temperature dependences that are consistent
with microscopic calculations.

However, there are remaining discrepancies.
For example, while the experimental and theoretical maximum IQEs can be made
to roughly agree, the current densities where they occur are overestimated
by the model. Several mechanisms may be responsible, e.g., effects arising
from doping profile, presence of carrier blocking layers and interface
irregularities \cite{schubert,choi}. These effects are ignored in the present study,
but may be readily incorporated into the model. Also excluded is a more
detailed description of defect recombination, e.g. as proposed for
describing the experimental data \cite{hader1,hader3}. More serious may is
the neglect of many-body effects, such as treated in Refs \cite{hader2} and 
\cite{kira}. There is also room for refinement in treating the connection
between the band-structure and population-dynamics aspects.
Carrying out these improvements increases computational demands, and
their implementation can be greatly facilitated by having more
experimental data and better knowledge details on the experimental
configurations. \bigskip

\textbf{Acknowledgement\smallskip }

This work is performed at Sandia's Solid-State Lighting Science Center, an
Energy Frontier Research Center (EFRC) funded by the U. S. Department of
Energy, Office of Science, Office of Basic Energy Sciences. The author
thanks helpful discussions with J. Wierer, the hospitality of the Technical
 University Berlin and travel support provided by SFB787.

\newpage\noindent{\large \textbf {Figure captions}}

\bigskip
\noindent Fig. 1. IQE versus current density for LED with single In$_{0.37}$Ga$_{0.63}
$N quantum well and temperatures $T_{L}=100$ to $400K$ at $50K$ intervals.
The arrows for the $T_{L}=200K$ IQE curve indicate the current densities 
for Figs. 3(b) and 3(c).

\bigskip
\noindent Fig. 2. Values of (a) SRH coefficient, (b) Auger coefficient and (c)
carrier-phonon scattering rate used in producing the curves in Fig. 1.

\bigskip
\noindent Fig. 3. (a) Spontaneous emission contributions from QW and barrier states
(solid and dashed curves, respectively) versus current density for $T=200K$
curve in Fig. 1. Solid curves in (b) and (c) are absolute square of envelope
functions at zone center ($k_{\bot }=k=0$) for electrons and holes at
current densities indicated by arrows in Fig. 3 (a). The x-axis is along the
growth direction. Each curve is displaced according to its energy for
clarity. Envelope functions belonging to QW and barrier states are indicated
by black and grey curves, respectively. The black dashed lines plot the
confinement potentials.

\bigskip
\noindent Fig. 4. Contributions to $T_{L}=200K$ IQE curve in Fig. 1.

\bigskip
\noindent Fig. 5. (a) IQE and (b) plasma temperature versus current density. In both
plots, the lattice temperature is $T_{L}=200K$ and Auger coefficients are $%
C=2.3\times 10^{-31}$ and $10^{-34}cm^{6}s^{-1}$ (solid and dashed curves,
respectively).

\bigskip
\noindent Fig. 6. IQE versus current density for LED with single 3nm In$_{0.20}$Ga$%
_{0.80}$N QW and temperatures $T=100K$ to $400K$ at $50K$ intervals. The SRH
and Auger coefficients, as well as the carrier-phonon scattering rates are
same as those in Fig. 1 for the In$_{0.37}$Ga$_{0.63}$N structure.

\bigskip
\noindent  Fig. 7. IQE versus current density for LED with (a) 3nm In$_{0.20}$Ga$%
_{0.80}$N QW and (b) 2nm In$_{0.37}$Ga$_{0.63}$N QW. The lattice\
temperatures are $T=100K$ to $300K$ at $50K$ intervals. The SRH and Auger
coefficients, as well as the carrier-phonon scattering rates are plotted in
Fig. 8.

\bigskip
\noindent Fig. 8. Values of (a) SRH coefficient, (b) Auger coefficient and (c)
carrier-phonon scattering rate used in producing the IQE curves in Fig. 7.


\begin{thebibliography}{99}
\bibitem{krames} M.R. Krames, O.B. Shchekin, R. Mueller-Mach, G.O. Mueller,
L. Zhou, G. Harbers, and M. G. Craford, \textquotedblleft Status and future
of high-power light-emitting diodes for solid-state
lighting,\textquotedblright\ J. Display Technology \textbf{3}, 160-175
(2007).

\bibitem{shen} Y.C. Shen, G.O. M\"{u}ller, S. Watanabe, N.F. Gardner, A.
Munkholm, and M.R. Krames, \textquotedblleft Auger recombination in InGaN
measured by photoluminescence,\textquotedblright\ Appl. Phys. Lett. \textbf{%
91}, 141101-141101 (2007).

\bibitem{hader2} J. Hader, J.V. Moloney, B. Pasenow, S.W. Koch, M. Sabathil,
N. Linder, and S. Lutgen, \textquotedblleft On the important of radiative
and Auger losses in GaN-based quantum wells,\textquotedblright\ Appl. Phys.
Lett. \textbf{92}, 261103-261105 (2008).

\bibitem{dellaney} K.T. Dellaney, P. Rinke, and C.G. Van de Walle,
\textquotedblleft Auger recombination rates in nitrides from first
principles,\textquotedblright\ Appl. Phys. Lett. \textbf{94}, 191109-191111
(2009).

\bibitem{schubert} M. H. Kim, M. F. Schubert, Q. Dai, J. K. Kim, E. F.
Schubert, J. Piprek, and Y. Park, \textquotedblleft Origin of efficiency
droop in GaN-based light-emitting diodes,\textquotedblright\ Appl. Phys.
Lett. \textbf{91}, 183507-183510 (2007).

\bibitem{efremov} A. A. Efremov, N. I. Bochkareva, R. I. Gorbunov, D. A.
Larinvovich, Yu. T. Rebane, D. V. Tarkhin, and Yu. G. Shreter,
\textquotedblleft Effect of the joule heating on the quantum efficiency and
choice of thermal conditions for high-power blue InGaN/GaN
LEDs,\textquotedblright\ Semiconductors \textbf{40}, 605 (2006).

\bibitem{chichibu} S. F. Chichibu, T. Azuhata, M. Sugiyama, T. Kitamura, Y.
Ishida, H. Okumurac, H. Nakanishi, T. Sota, and T. Mukai, \textquotedblleft
Optical and structural studies in InGaN quantum well structure laser
diodes,\textquotedblright\ J. Vac, Sci. Technol. B \textbf{19}, 2177 (2001).

\bibitem{smowton} I. A. Pope, P.\ M. Smowton, P. Blood, and J. D. Thompson,
\textquotedblleft Carrier leakage in InGaN quantum well light-emitting
diodes emitting at 480nm,\textquotedblright\ Appl. Phys. Lett. \textbf{82},
2755-2757 (2003).

\bibitem{hader1} J. Hader, J.V. Moloney, and S.W. Koch, \textquotedblleft
Density-activated defect recombination as a possible explanation for the
efficiency droop in GaN-based diodes,\textquotedblright\ Appl. Phys. Lett. 
\textbf{96}, 221106-221108 (2010).




\bibitem{chow1} W. W. Chow, M. H. Crawford, J. Y. Tsao, and M. Kneissl,
\textquotedblleft Internal efficiency of InGaN light-emitting diodes: Beyond
a quasiequilibrium model,\textquotedblright\ Appl. Phys. Lett. \textbf{97},
121105-121107 (2010).

\bibitem{chow2} W. W. Chow, `Modeling excitation-dependent bandstructure
effects on InGaN light-emitting diode efficiency,' Optics Express \textbf{19}%
, 21818-21831, 2011.

\bibitem{shur} A. Bykhovshi, B. Gelmonst, and M. Shur, \textquotedblleft The
influence of the strain-induced electric field on the charge distribution in
GaN-AlN-GaN structure,\textquotedblright\ J. Appl. Phys. \textbf{74},
6734-6739 (1993).

\bibitem{hangleiter} J. S. Im, H. Kollmer, J. Off, A. Sohmer, F. Scholz, and
A. Hangleiter, \textquotedblleft Reduction of oscillator strength due to
piezoelectric fields in GaN/AlGaN quantum wells,\textquotedblright\ Phys.
Rev. B \textbf{57}, R9435-R9438 (1998).

\bibitem{kira} W. Chow, M. Kira and S. W. Koch, `Microscopic theory of
optical nonlinearities and spontaneous emission in group-III nitride quantum
wells,' Phys. Rev. B. \textbf{60}, 1947-1952 (1999).

\bibitem{ryu} H.-Y Ryu, H.-S. Kim, and J.-I. Shim, \textquotedblleft Rate
equation analysis of efficiency droop in InGaN light-emitting
diodes,\textquotedblright\ Appl. Phys. Lett. \textbf{95}, 081114-081117
(2009).

\bibitem{hader3} J. Hader, J.V. Moloney and S.W. Koch, \textquotedblleft
Temperature-dependence of the internal efficiency droop in GaN-based
diodes,\textquotedblright\ Appl. Phys. Lett. \textbf{99}, 181127-3 (2011).

\bibitem{jaynes} E. Jaynes, and F. Cummings, \textquotedblleft Comparison of
quantum and semiclassical radiation theories with application to the beam
maser,\textquotedblright\ Proc. IEEE \textbf{51}, 89-109 (1963).

\bibitem{waldmueller} I. Waldmueller, W. W. Chow, M. C. Wanke, and E. W.
Young, \textquotedblleft Non-equilibrium many-body theory of intersubband
lasers,\textquotedblright\ IEEE J. Quantum Electron. \textbf{42}, 292-301
(2006).

\bibitem{chuang} S. L. Chuang, and C. S. Chang, \textquotedblleft $k\cdot p$
method for strained wurtzite semiconductors,\textquotedblright\ Phys. Rev. B 
\textbf{54}, 2491-2504 (1996).

\bibitem{jenkins} S. J. Jenkins, g. P. Srivastava, and J. C. Inkson,
\textquotedblleft Simple approach to self-energy corrections in
semiconductors and insulators,\textquotedblright\ Phys. Rev. B \textbf{48},
4388 (1993).

\bibitem{wright} A. F. Wright, and J. S. Nelson, \textquotedblleft
Consistent structural properties for AlN, GaN, and InN,\textquotedblright\
Phys. Rev. B \textbf{51}, 7866-7869 (1995).

\bibitem{wei} S. H. Wei, and A. Zunger, \textquotedblleft Valence band
splittings and band offsets of AlN, GaN, and InN,\textquotedblright\ Appl.
Phys. Lett. \textbf{69}, 2719-2711 (1996).

\bibitem{ambacher} O. Ambacher, \textquotedblleft Growth and applications of
Group III-nitrides,\textquotedblright\ J. Phys. D: Appl. Phys. \textbf{31},
2653-2710 (1998).

\bibitem{expt1} A. Laubsch, M. Sabathil, W. Bergbauer, M. Strassburg, H.
Lugauer, M. Peter, S. Lutgen, N. Linder, K. Strebubel, J. Wagner, J. Hader,
J. V. Moloney and S.W. Koch, \textquotedblleft On the origin of IQE-'droop'
in InGaN LEDs,\textquotedblright\ Phys. Status Solidi C 6, S913-S916 (2009).

\bibitem{expt2} K. Fujiwara, H. Jimi and K. Kaneda, \textquotedblleft
Temperature-dependent droop of electroluminescence efficiency in blue
(In,Ga)N quantum-well diodes,\textquotedblright\ Phys. Status Solidi C 6,
S814-S817 (2009).

\bibitem{knorr} W. W. Chow, A. Knorr, and S. W. Koch, \textquotedblleft
Theory of laser gain in group-III nitrides,\textquotedblright\ Appl. Phys.
Lett. \textbf{67}, 754-756 (1995).

\bibitem{choi} S. Choi, H. J. Kim, S.-S. Kim, J. Liu, J. Kim, J.-H. Ryou, R.
D. Dupuis, A. M. Fishcer, and F. A. Ponce, \textquotedblleft Improvement of
peak quantum efficiency and efficiency droop in III-nitride visible
light-emitting diodes with an InAlN electron-blocking
layer,\textquotedblright\ Appl. Phys. Lett. \textbf{96}, 221105-221107
(2010).
\end{thebibliography}
\end{document}